\documentstyle[twoside,fleqn,epsfig]{actaps}

\begin{document}
 \newcommand{\beq}{\begin{equation}}\newcommand{\eeq}{\end{equation}}
 \newcommand{\barr}{\begin{eqnarray}}\newcommand{\earr}{\end{eqnarray}}
\newcommand{\andy}[1]{ }

 \def\tltl{\widetilde}
 \def\bmn{\mbox{\boldmath $n$}} \def\bmA{\mbox{\boldmath $A$}}
 \def\bmB{\mbox{\boldmath $B$}} \def\bmb{\mbox{\boldmath $b$}}
 \def\bmsigma{\mbox{\boldmath $\sigma$}}
 \def\bmsigman{\mbox{\boldmath $\sigma$}\cdot\mbox{\boldmath $n$}}
 \def\bmsigmab{\mbox{\boldmath $\sigma$}\cdot\mbox{\boldmath $b$}}
 \def\bmsigmaA{\mbox{\boldmath $\sigma$}\cdot\mbox{\boldmath $A$}}
 \def\bmsigmaB{\mbox{\boldmath $\sigma$}\cdot\mbox{\boldmath $B$}}
 \def\ch{\mbox{ch}}
 \def\sh{\mbox{sh}}
 \def\eqn#1{Eq.\ (\ref{eq:#1})}
 \def\coltwovector#1#2{\left({#1\atop#2}\right)}
 \def\up{\coltwovector10}
 \def\down{\coltwovector01}
 \newcommand{\bm}[1]{\mbox{\boldmath $#1$}}
 \newcommand{\bmsub}[1]{\mbox{\boldmath\scriptsize $#1$}}
 \newcommand{\ket}[1]{| #1 \rangle}
 \newcommand{\bra}[1]{\langle #1 |}

\headings{1}{8}
\def\authorlist{P.\ Facchi, A. Mariano, S.\ Pascazio}
\def\shorttitle{Neutron coherence}

\title{\uppercase{Wigner function and coherence properties of cold and thermal neutrons}}

\author{P.\ Facchi$^{a b}$\email{paolo.facchi@ba.infn.it},
A.\ Mariano$^{a}$\email{angelo.mariano@ba.infn.it}, S.\
Pascazio$^{a b}$\email{saverio.pascazio@ba.infn.it}} { $(a)$
Dipartimento di Fisica, Universit\`a di Bari \\ $(b)$ Istituto
Nazionale di Fisica Nucleare, Sezione di Bari \\
 I-70126  Bari, Italy
}

\day{30 April 1999}  

\abstract{%
We analyze the coherence properties of a cold or a thermal neutron
by utilizing the Wigner quasidistribution function. We look in
particular at a recent experiment performed by Badurek {\em et
al.}, in which a polarized neutron crosses a magnetic field that is
orthogonal to its spin, producing highly non-classical states. The
quantal coherence is extremely sensitive to the field fluctuation
at high neutron momenta. A ``decoherence parameter" is introduced
in order to get quantitative estimates of the losses of coherence.
}

\medskip


 \setcounter{equation}{0}
 \section{Introduction }
 \label{sec-introd}
 \andy{intro}

Highly non-classical, Schr\"odinger-cat-like neutron states can be
produced by coherently superposing different spin states in an
interferometer and with neutron spin echo \cite{RS,RSP}. We analyze
here an interesting recent experiment \cite{BRSW} in which a
polarized neutron crosses a magnetic field that is orthogonal to
its spin, producing Schr\"odinger-cat-like states. Our main purpose
is to investigate the decoherence effects that arise when the
fluctuations of the magnetic field are considered.

\section{Squeezing and squashing}
\label{sec-sqesqa}
\andy{sqesqa}

Let us start by looking at the coherence properties of a neutron
wave packet and concentrate our attention on the losses of
coherence provoked by a fluctuating magnetic field. To this end, we
introduce the Wigner quasidistribution function
\andy{wigdef}
\beq\label{eq:wigdef}
W(x,k) = \frac{1}{2\pi}\int d\xi\; e^{-ik\xi}
\psi \left(x+\frac{\xi}{2} \right)
\psi^*\left(x-\frac{\xi}{2} \right) ,
\eeq
where $x$ is position, $p=\hbar k$ momentum and $\psi$ the wave
function of the neutron in the apparatus. The Wigner function is
normalized to one and its marginals represent the position and
momentum probability distributions
\andy{margxp}
\beq
\int dx\; dk\; W(x,k) =1; \qquad
P(x) = \int dk\; W(x,k), \quad P(k) = \int dx\; W(x,k).
\label{eq:margxp}
\eeq
We shall work in one dimension. We assume that the neutron wave
function is well approximated by a Gaussian
\andy{gauss,gaussinv}
\barr
\psi(x) &=& \frac{1}{(2\pi\delta^2)^{1/4}} \exp
\left[-\frac{(x-x_0)^2}{4\delta^2} + i k_0 x\right] ,
\label{eq:gauss} \\
\phi(k) &=& \frac{1}{(2\pi\delta_k^2)^{1/4}} \exp
\left[-\frac{(k-k_0)^2}{4\delta_k^2} - i(k-k_0)x_0\right]
\nonumber \\
 &=&
\left(\frac{2\delta^2}{\pi} \right)^{1/4}
\exp \left[-\delta^2(k-k_0)^2 - i(k-k_0)x_0\right],
\label{eq:gaussinv}
\earr
where $\delta$ is the spatial spread of the wave packet, $\delta_k
\delta= 1/2$, $x_0$ is the initial average position of the neutron
and $p_0=\hbar k_0$ its average momentum. The two functions above
are related by a Fourier transformation and are both normalized to
one. Normalization will play an important role in our analysis and
will never be neglected.

The Wigner function for the state
(\ref{eq:gauss})-(\ref{eq:gaussinv}) is readily calculated
\andy{wiggauss}
\beq\label{eq:wiggauss}
W(x,k) = \frac{1}{\pi}
\exp \left[-\frac{(x-x_0)^2}{2\delta^2}\right]
\exp \left[-2\delta^2(k-k_0)^2\right]
\eeq
and turns out to be a positive function. In the language of quantum
optics \cite{QOpt}, we shall say that the neutron is prepared in a
coherent state if $\delta =
\delta_k= 1/\sqrt{2}$ and in a squeezed state
if $\delta \neq \delta_k$. An illustrative example is
given in Figure 1.
 \begin{figure}
\centerline{\epsfig{file=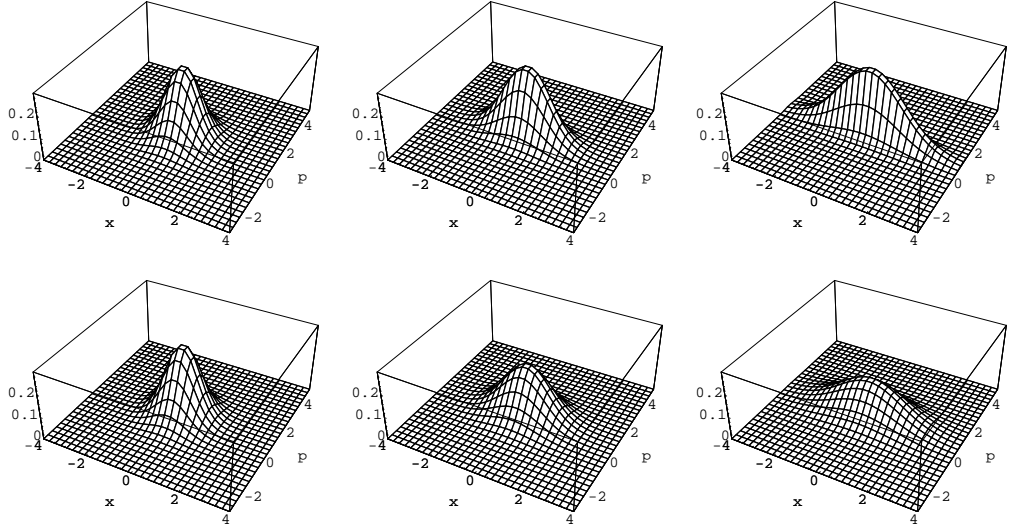, width=\textwidth}}
 \caption{Fig. 1. Above, Wigner function
(\ref{eq:wiggauss}), for $x_0=0, k_0=1.7 \cdot 10^{10}$m$^{-1}$;
$x$ is in units $10^{-10}$m and $p=k$ in units $10^{10}$m$^{-1}$.
From left to right: $\delta= 1/\sqrt{2}$ (coherent state),
$\delta=1$ (squeezed state) and $\delta =
\sqrt{2}$ (a more squeezed state); the uncertainty principle always
reads $\delta_k\delta=1/2$ (minimum uncertainty states). Below,
Wigner function in (\ref{eq:wigm}) for the same values of $x_0,k_0$
and $\Delta_0=0$. From left to right:
$\sigma=0,1/\sqrt{2},\sqrt{3/2}$; the uncertainty principle yields
$\delta_k\delta'=1/2,1/\sqrt{2},1$} (squashing).
 \end{figure}

Consider now a polarized neutron that crosses a constant magnetic
field, parallel to its spin, of intensity $B$ and contained in a
region of length $L$. Since the total energy is conserved, the
kinetic energy of the neutron in the field changes by $\Delta E =
\mu B >0$, where $-\mu$ is the neutron magnetic moment.
This implies a change in average momentum $\Delta k= m\mu B
/\hbar^2 k_0$ and an additional shift of the neutron phase
proportional to $\Delta
\equiv L\Delta k/k_0$. The resulting effect on the Wigner function is
$W(x,k)\to W(x-\Delta,k)$.

Assume now that the intensity of the $B$-field fluctuates around
its average $B_0$ according to a Gaussian law. This fluctuation is
reflected in a fluctuation of the quantity $\Delta$ according to
the distribution law
\andy{probdel}
\beq\label{eq:probdel}
w(\Delta) = \frac{1}{\sqrt{2\pi\sigma^2}}
\exp \left[-\frac{(\Delta-\Delta_0)^2}{2\sigma^2}\right],
\eeq
where $\sigma$ is the standard deviation. The ratio
$\sigma/\Delta_0$ is simply equal to the ratio $\delta B/B_0$,
$\delta B$ being the standard deviation of the fluctuating
$B$-field. The average Wigner function, when the neutron has
crossed the whole $B$ region of lenght $L$, represents a
``squashed" state, that has partially lost its quantum coherence:
\andy{wigm}
\beq\label{eq:wigm}
W_{\rm m}(x,k)= \int d\Delta\; w(\Delta)\; W(x-\Delta,k).
\eeq
This function is represented in Figure~1 for $\Delta_0=0$
(vanishing average magnetic field) and increasing values of
$\sigma$. The above Wigner function can be calculated explicitly,
but its expression is a bit cumbersome; however, its marginals
(\ref{eq:margxp}) are simple:
\andy{marg1,marg2}
\barr
P(x) &=&
\frac{1}{\sqrt{2\pi(\delta^2+\sigma^2)}} \exp
\left[-\frac{(x-x_0-\Delta_0)^2}{2(\delta^2+\sigma^2)}\right] ,
\label{eq:marg1} \\
P(k) &=&
\sqrt{\frac{2\delta^2}{\pi} }
\exp \left[-2\delta^2(k-k_0)^2\right].
\label{eq:marg2}
\earr
Notice that the momentum distribution (\ref{eq:marg2}) is unaltered
[$|\phi(k)|^2$ in (\ref{eq:gaussinv})]: obviously, the energy of
the neutron does not change. Observe the additional spread in
position $\delta'=(\delta^2+\sigma^2)^{1/2}$ and notice that the
Wigner function and its marginals are always normalized to one. The
uncertainty principle yields $\delta_k\delta'
=\frac{1}{2}
\sqrt{1+\sigma^2/\delta^2} > 1/2$.

\section{Schr\"odinger-cat states in a fluctuating magnetic field}
\label{sec-noise}
\andy{noise}
Let us now look in more detail at the experiment \cite{BRSW}. A
polarized ($+y$) neutron enters a magnetic field, perpendicular to
its spin, of intensity $B_0=0.28$mT, confined in a region of length
$L=57$cm. Due to Zeeman splitting, the two neutron spin states
travel with different speeds in the field. The average neutron
wavenumber is $k_0=1.7 \cdot 10^{10}$m$^{-1}$ and its coherence
length (defined by a chopper) is $\delta=1.1\cdot 10^{-10}$m. By
travelling in the magnetic field, the two neutron spin states are
separated by a distance $\Delta_0=2 m\mu B_0/\hbar^2 k_0= 16.1
\cdot 10^{-10}$m, one order of magnitude larger than $\delta$
(notice the factor 2, absent in the definition of the previous
section). Observe that the neutron wave packet itself has a natural
spread $\delta_t=(\delta^2+(\hbar t/2m\delta)^2)^{1/2} \simeq 15$cm
(due to its free evolution for a time $t\simeq  mL/\hbar k_0$);
however, we shall neglect this additional effect, because it is
irrelevant for the loss of quantum coherence.

After the neutron has crossed the $B$-field only the $+y$
spin-component is observed and its Wigner function is readily
computed
\barr
& & W(x,k)=\frac{1}{4\pi}\exp[-2\delta^2(k-k_0)^2]\nonumber\\ & &
\times\left[\exp\left(-\frac{\left(x-\frac{\Delta}{2}\right)^2}{2\delta^2}\right)
+\exp\left(-\frac{\left(x+\frac{\Delta}{2}\right)^2}{2\delta^2}\right)
+2
\exp\left(-\frac{x^2}{2\delta^2}\right)\cos(k\Delta)\right].
\nonumber\\
\earr
Notice that for $\Delta=0$ (no $B$-field) one obtains
(\ref{eq:wiggauss}). Our interest is to investigate the loss of
quantum coherence if the intensity of the $B$-field fluctuates,
like in the previous section, yielding a random shift according to
the law (\ref{eq:probdel}). In such a case, the average Wigner
function reads
\andy{Wm2}
\barr
& & W_{\rm m}(x,k)=\int d\Delta\;w(\Delta)\;W(x,k) =
\frac{1}{4\pi}\exp[-2\delta^2(k-k_0)^2] \nonumber \\
 & & \times \left[\sqrt{\frac{\delta^2}{\delta^2+\frac{\sigma^2}{4}}}
\exp\left(-\frac{\left(x-\frac{\Delta_0}{2}\right)^2}
{2\left(\delta^2+\frac{\sigma^2}{4}\right)}\right)
 +\sqrt{\frac{\delta^2}{\delta^2+\frac{\sigma^2}{4}}}
\exp\left(-\frac{\left(x+\frac{\Delta_0}{2}\right)^2}
{2\left(\delta^2+\frac{\sigma^2}{4}\right)}\right)
\right. \nonumber\\
& &\quad\quad +\left.
2\exp\left(-\frac{x^2}{2\delta^2}\right)
\exp\left(-\frac{\sigma^2 k^2}{2}\right)\cos(k\Delta_0)\right]
\label{eq:Wm2}
\earr
\begin{figure}
\centerline{\epsfig{file=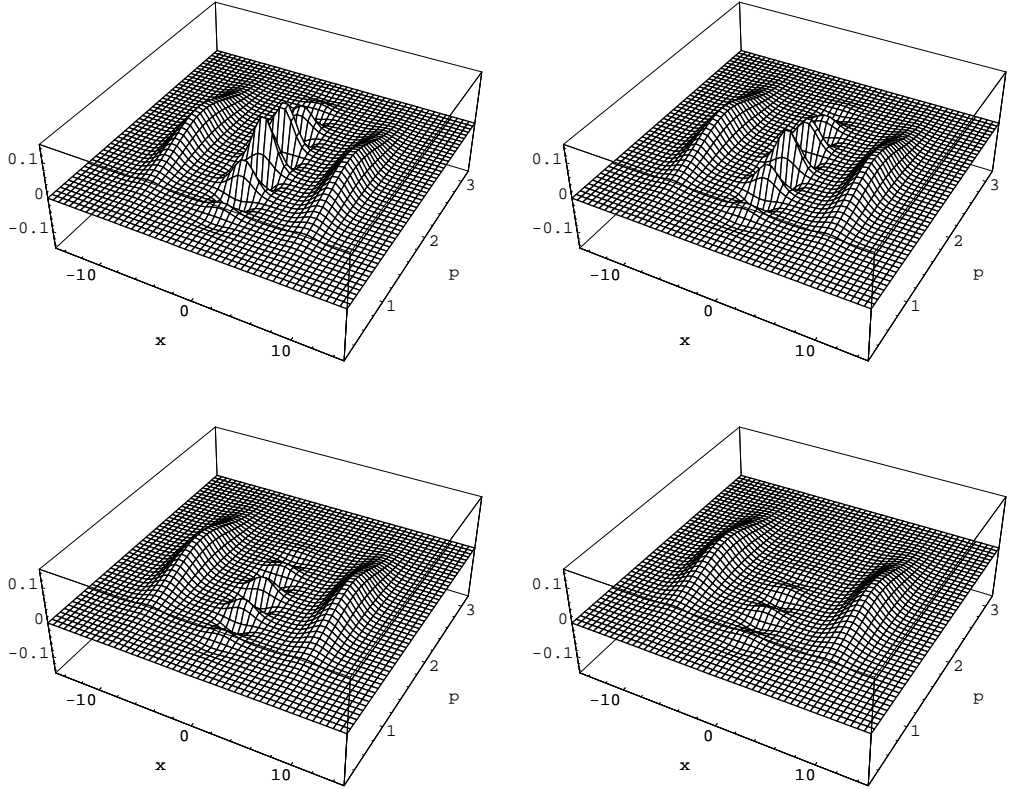, width=\textwidth}}
 \caption{Fig. 2. Wigner function (\ref{eq:Wm2})
for $\Delta_0=0, k_0=1.7 \cdot 10^{10}$m$^{-1},\delta=1.1 \cdot
10^{-10}$m; $x$ is measured in units $10^{-10}$m and $p=k$ in units
$10^{10}$m$^{-1}$. From above left to bottom right,
$\sigma=0,0.5,1.0,1.5\cdot10^{-10}$m. Notice the asymmetry of the
``wiggles" around $k_0$ and their fragility  at high values of
momentum. Observe also the slight ``squashing" of the Gaussian
components at high values of $\sigma$.
 }
 \end{figure}
and the momentum distribution function yields
\andy{Wm2pk}
\barr
P(k)= \sqrt{\frac{\delta^2}{2\pi}}\exp[-2\delta^2(k-k_0)^2]
\left[ 1 + \exp\left(-\frac{\sigma^2 k^2}{2}\right)\cos(k\Delta_0)\right].
\label{eq:Wm2pk}
\earr
Notice also that, since only the $+y$-component of the neutron spin
is observed, the normalization reads
\barr
N &=& \int dx\; dk\; W_{\rm m}(x,k) \nonumber\\ &=& \frac{1}{2}
\left[1 +\sqrt{\frac{\delta^2}{\delta^2+\frac{\sigma^2}{4}}}
\exp\left(-\frac{\Delta_0+4\delta^2\sigma^2 k_0^2}
{8\left(\delta^2+\frac{\sigma^2}{4}\right)}\right)
\cos\left(\frac{\delta^2}{\delta^2+\frac{\sigma^2}{4}}
 k_0\Delta_0\right)\right].
\earr
Obviously, $N=1$ when no magnetic field is present
($\sigma=\Delta_0=0$). The Wigner function (\ref{eq:Wm2}) is
plotted in Figure 2 for some values of $\sigma$. The off-diagonal
part of the Wigner function (``trustee" of the interference
effects) is very fragile at high values of momentum. This was
already stressed in \cite{RSP,BRSW} and is apparent in the
structure of the marginal distribution (\ref{eq:Wm2pk}): the term
$\exp (-\sigma^2 k^2/2)$ strongly suppresses the interference
effects at high $k$'s.

\section{Decoherence parameter}
\label{sec-dec}
\andy{dec}

One can give a quantitative estimate of the loss of quantum
coherence by introducing a ``decoherence parameter," in the same
spirit of Refs.\ \cite{MHS}. To this end, remember that the Wigner
function can be expressed in terms of the density matrix $\rho$ as
\andy{wigrho}
\beq\label{eq:wigrho}
W(x,k) = \frac{1}{2\pi}\int d\xi\; e^{-ik\xi}
\langle x+\xi/2|\rho|x-\xi/2\rangle ,
\eeq
and that $\mbox{Tr}(\rho^2) =\mbox{Tr}\rho =1$ for a pure state,
while $\mbox{Tr}(\rho^2) < \mbox{Tr} \rho=1$ for a mixture. Define
therefore the {\em decoherence parameter}
\andy{decpar}
\beq\label{eq:decpar}
\varepsilon (\sigma) =1-\frac{\mbox{Tr}(\rho^2)}{(\mbox{Tr}\rho)^2} =
1 - \frac{2\pi \int dx\; dk\; W_{\rm m}(x,k)^2} {\left(\int dx\;
dk\; W_{\rm m}(x,k)\right)^2} .
\eeq
This quantity is expected to vanish for $\sigma=0$ (no fluctuation
of the $B$-field and quantum coherence perfectly preserved) and to
become unity when $\sigma \to \infty$ (large fluctuations of the
$B$-field and quantum coherence completely lost). Figure 3 confirms
these expectations, that can also be proven analitically from
(\ref{eq:Wm2}).
 \begin{figure}
\centerline{\epsfig{file=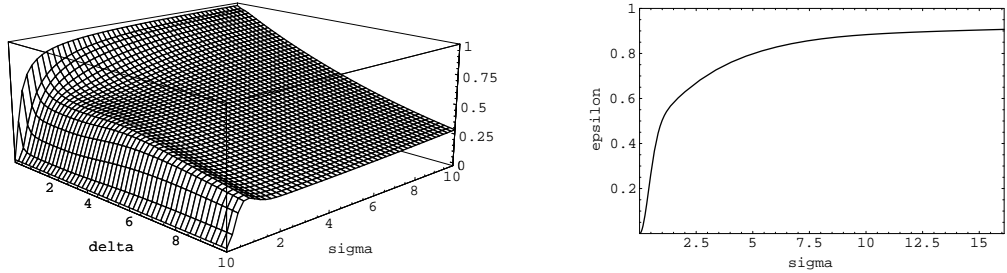, width=\textwidth}}
 \caption{Fig. 3. Decoherence parameter.
 Left: $\epsilon$ as a function of $\delta$ and $\sigma$ (both in units
 $10^{-10}$m). Notice the peculiar behavior when $\delta>3$ and $1<\sigma<2$.
Right: $\epsilon$ vs $\sigma$ (in $10^{-10}$m) for $\delta=1.1\cdot
10^{-10}$m (experimental value in \cite{BRSW}).
 }
 \end{figure}
In Ref.\ \cite{BRSW}, $\delta=1.1 \cdot 10^{-10}$m and $\sigma$ is
(presumably) very small, being the intensity of the $B$ field
controlled with high accuracy. It is remarkable that the
decoherence parameter is not a monotonic function of the noise
$\sigma$, when $\delta>3\cdot10^{-10}$m and
$1\cdot10^{-10}$m$<\sigma<2\cdot10^{-10}$m. This may be due to our
very definition (\ref{eq:decpar}) or to some physical effect we do
not yet understand.

\medskip
\noindent {\bf Acknowledgments:}
We thank G.\ Badurek, H. Rauch and M. Suda for many useful
discussions. This work was partially supported by the TMR Network
``Perfect Crystal Neutron Optics" (ERB-FMRX-CT96-0057) of the
European Union.


 \end{document}